\documentclass[journal=ancac3,manuscript=article]{achemso}

\usepackage{amsmath,amssymb,amsfonts}
\usepackage{bm}
\usepackage{color}
\usepackage{graphicx}
\usepackage{float}
\usepackage{verbatim}
\usepackage[english]{babel}
\usepackage{hyperref}
\usepackage{natbib}
\usepackage{subfigure}
\usepackage{amstext, mathrsfs, textcomp}
\usepackage{multirow}
\usepackage{enumerate}
\usepackage{epstopdf}
\usepackage{tabularx}

\makeatletter

\newcommand{\VB}[1]{\mathbf{#1}} 

\title{Coherent control of topological states \\ in an integrated waveguide lattice}

\author{Alexey Mikhin}
\affiliation{School of Physics and Engineering, ITMO University, Saint Petersburg 197101, Russia}
\alsoaffiliation{Authors who contributed equally to this work}

\author{Viktoriia Rutckaia}
\affiliation{Photonics Initiative, Advanced Science Research Center, The City University of New York, New York 10031, USA}
\alsoaffiliation{Centre for Innovation Competence SiLi-nano, Martin-Luther-University, Halle-Wittenberg, 06120 Halle (Saale), Germany}
\alsoaffiliation{Authors who contributed equally to this work}

\author{Roman S. Savelev}
\affiliation{School of Physics and Engineering, ITMO University, Saint Petersburg 197101, Russia}
\alsoaffiliation{Authors who contributed equally to this work}

\author{Ivan S. Sinev}
\affiliation{School of Physics and Engineering, ITMO University, Saint Petersburg 197101, Russia}

\author{Andrea Al\`u}
\affiliation{Photonics Initiative, Advanced Science Research Center, The City University of New York, New York 10031, USA}
\alsoaffiliation{Physics Program, Graduate Center, The City University of New York, New York 10016, USA}

\author{Maxim A. Gorlach}
\affiliation{School of Physics and Engineering, ITMO University, Saint Petersburg 197101, Russia}
\email{m.gorlach@metalab.ifmo.ru}


\begin{document}

\maketitle

\begin{abstract}
Topological photonics holds the promise for enhanced robustness of light localization and propagation enabled by the global symmetries of the system. While traditional designs of topological structures rely on lattice symmetries, there is an alternative strategy based on accidentally degenerate modes of the individual meta-atoms. Using this concept, we experimentally realize topological edge state in an integrated optical nanostructure based on the array of silicon nano-waveguides, each hosting a pair of  degenerate modes at telecom wavelengths. Exploiting the hybrid nature of the topological mode formed by the superposition of waveguide modes with different symmetry, we implement coherent control of the topological edge state by adjusting the phase between the degenerate modes and demonstrating selective excitation of bulk or edge states. The resulting field distribution is imaged via third harmonic generation allowing us to quantify the localization of topological modes as a function of the relative phase of the excitations. Our results highlight the impact of engineered accidental degeneracies on the formation of topological phases, extending the opportunities stemming from topological nanophotonic systems.
\end{abstract}

\hfill \break

\section{Introduction}\label{sec:intro}

Topological photonics provides a promising avenue to manipulate light in engineered nanostructures by creating disorder-robust edge or interface states immune to backscattering at sharp bends and defects~\cite{Lu2014,Lu2016,Ozawa_RMP}. The first approaches to tailor such states have been relying on broken time-reversal symmetry by applying an external magnetic bias in magneto-optical materials~\cite{Raghu-PRA-2008,Haldane-PRL-2008,Wang-Soljacic}. This strategy however faces practical limitations, which have been inspiring a search for alternative time-reversal-invariant platforms~\cite{Rechtsman,Hafezi-2011,Hafezi:2013NatPhot,Khanikaev2013,Wu-Hu}, such as crystalline topological metamaterials~\cite{Wu-Hu,Yves2017,Barik,Li2018,Gorlach2018,Noh2018,Yang2018,Smirnova2019,Kuipers}, particularly appealing for its experimental accessibility.

In crystalline topological systems, the nontrivial topology of the bands and the associated edge or interface states originate due to the special choice of lattice symmetries that ensure topological degeneracies (e.g. Dirac points), as well as the opening of topological bandgaps for suitable lattice parameters. Prominent examples are one-dimensional (1D) zigzag arrays~\cite{Poddubny2014,Sinev2015,Slob2015,Solnyshkov,Kruk-Poddubny,StJean}, 2D breathing honeycomb~\cite{Wu-Hu,Barik} and breathing kagome~\cite{Ezawa2018,Xue2018,Ni2018} lattices. However, since lattice geometries are challenging to modify dynamically, the topological properties of such systems, e.g., the existence of topological states and their localization length, are difficult to control in real time.


To overcome this limitation, it has recently been suggested to utilize the accidental degeneracy of modes in the individual meta-atoms of a simple lattice~\cite{Aravena2020,Savelev2020}. In such case, the interference of the near fields of the degenerate modes can regulate the coupling between neighboring meta-atoms, while the detuning between these modes controls the topological transitions~\cite{Aravena2020,Savelev2020,Vicencio2021}. Experimental demonstrations of this strategy to create topological structures at the nanoscale have remained elusive so far.

\begin{figure}[t!]
\centering
\includegraphics[width=1\textwidth]{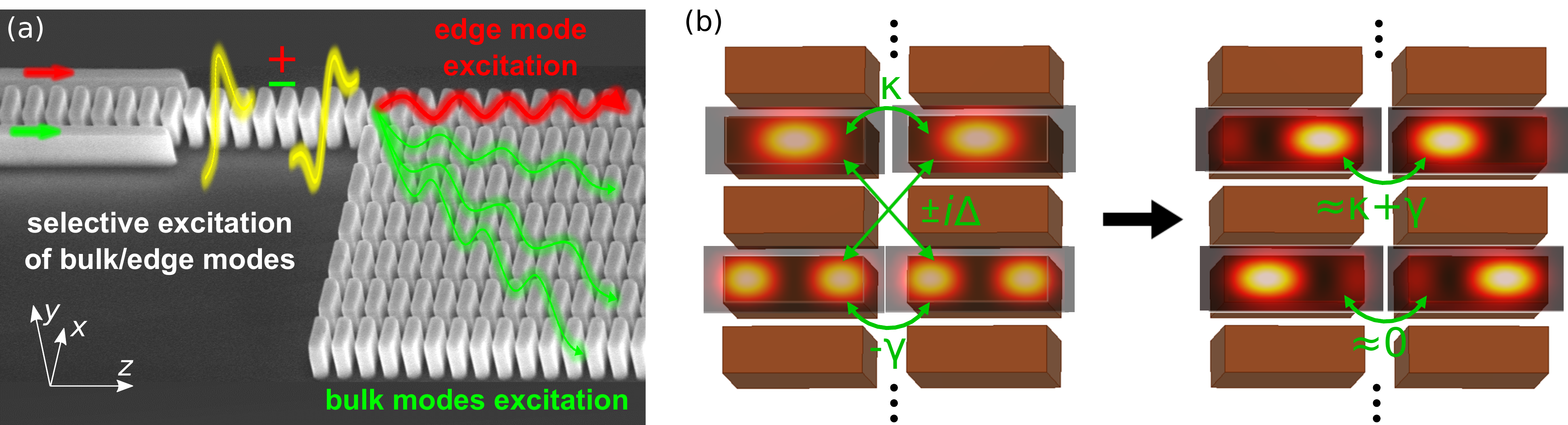}
\caption{(a) Scanning electron microscopy image (SEM) of the fabricated structure with schematic profiles of the near field distributions for two quasi-degenerate modes of the structured waveguide (yellow curves). Depending on the relative phase difference with which these modes are initiated through the directional couplers, either a topological edge mode (red curve) or bulk modes  (green curves) of the waveguide array are excited. (b) Schematic of the $|E_y|^2$ field distribution for the two modes of the neighboring waveguides. The coupling between them gives rise to the hybrid modes with asymmetric field profiles.}
\label{fig:scheme}
\end{figure}

In this Article, we implement this strategy experimentally realizing an integrated optical structure based on an array of multimode nanostructured waveguides that supports topological edge state in the telecom range.
By optimizing the design of the waveguides, we ensure that they host a pair of modes with the same propagation constants but different symmetry of the near field profiles [Figure~\ref{fig:scheme}a]. Such accidental degeneracy is crucial for the formation of a Floquet topological phase~\cite{Rechtsman} and the emergence of topological edge states. In contrast to the traditional implementations of topological physics, these modes have hybrid origin being a superposition of two waveguide modes with different symmetry of the near field. As we demonstrate, this feature enables a coherent control of the topological edge state.

Generally, the goal of coherent control phenomena is to tailor the relative phases of multiple excitation signals to control the response of the system in real-time~\cite{WarrenScience,Goswami2003,Rabitz2009}. In photonics, this is often achieved through the interference of excitations from multiple ports~\cite{Zhang2020,Kang2022}. In our case, coherent control is ensured by manipulating the relative phase between two waveguide modes, which allows us to switch between the excitation of edge and bulk modes of the lattice. To image the modes of the fabricated waveguide nanostructure, we excite it with short laser pulses and collect the third harmonic signal, which provides a sensitive tool for detection and localization.


\section{Design of the Waveguide Lattice}\label{sec:design}

The key ingredient necessary to obtain the topological features in our setup is the degeneracy of two modes with distinct symmetries of their near field distributions as sketched in Fig.~\ref{fig:scheme}b. If two degenerate modes $E_{y1}$ and $E_{y2}$ exhibit different parity of the field relative to reflection in $Oyz$ plane, their linear combinations become strongly asymmetric relative to the waveguide axis. This immediately affects the magnitude of the evanescent coupling between the adjacent waveguides [Fig.~\ref{fig:scheme}b].

In an array of waveguides, hybridization of their modes gives rise to bulk and edge states. If two degenerate modes of an edge waveguide are excited in phase, their linear combination $E_{y1}+E_{y2}$ will feature the hot-spot at the edge of the structure decoupled from the rest of the lattice. On the contrary, if the same modes are excited out of phase, their linear combination $E_{y1}-E_{y2}$ will feature the hot-spot on the other side of the waveguide ensuring good coupling to the bulk of the structure [Fig.~\ref{fig:scheme}b]. This reasoning suggests the emergence of an edge-localized state with a hybrid nature.

\begin{figure}[h!]
\centering
\includegraphics[width=1\textwidth]{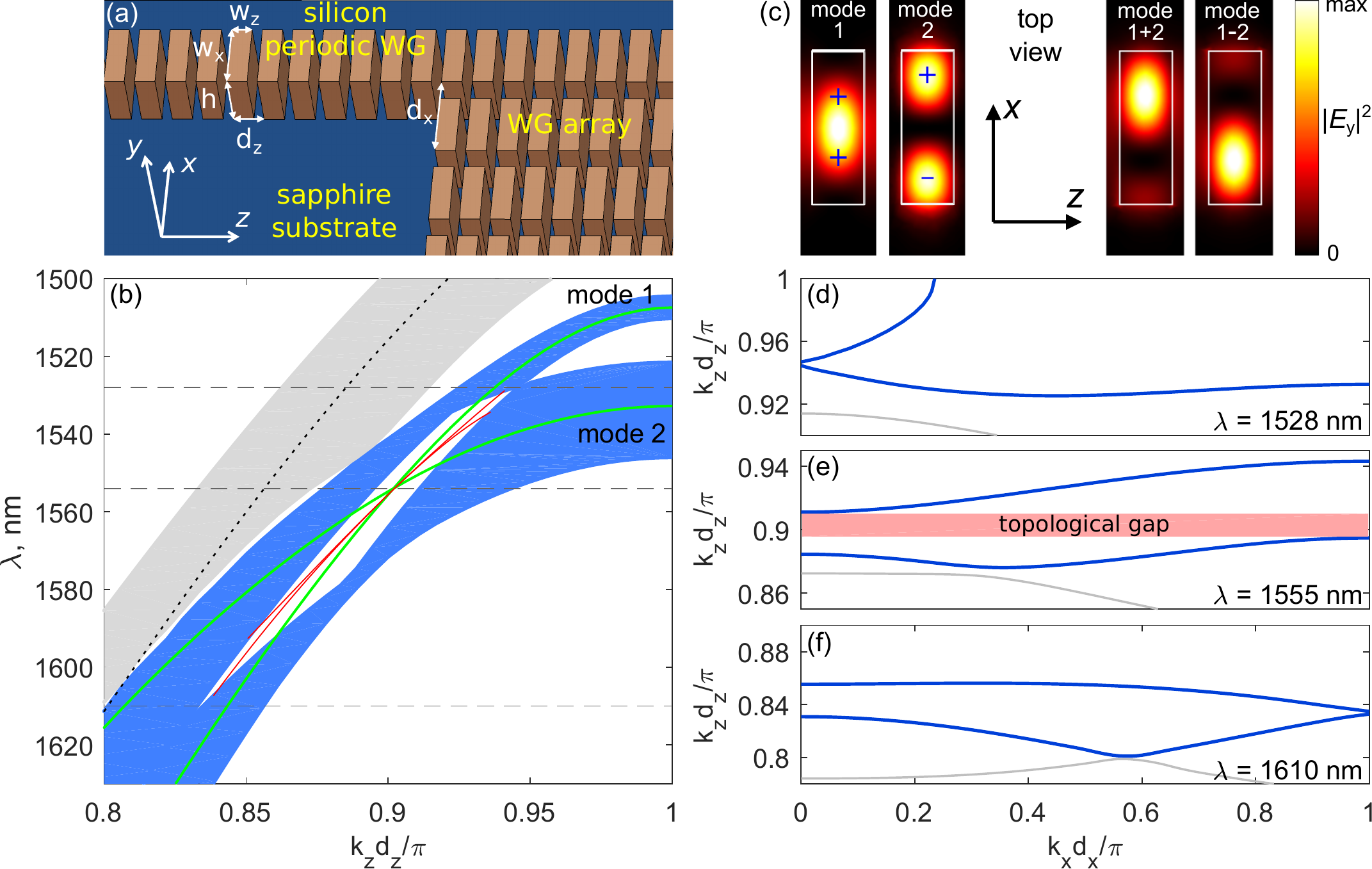}
\caption{(a) Schematic of a silicon waveguide array with extended edge waveguide on a sapphire substrate. The geometrical parameters of the structure are $h=600$~nm, $w_x=605$~nm, $w_z=205$~nm, $d_x=790$~nm, $d_z=300$~nm, respectively. (b) Simulated dispersion of the three modes of the individual waveguide with lowest frequency (solid green and black dotted curves), and the dispersion bands of the waveguide array (shaded areas). Two red curves show the dispersion of two nearly degenerate edge modes in the array of 7 waveguides. (c) Intensity distributions of the $y$ component of the electric field $|E_y|^2$ in the unit cell of the individual periodic waveguide. Left two panels: $|E_y|^2$ for hybrid HE$_{11}$ and quasi-TE$_{01}$ modes, respectively, at their intersection frequency, ``+'' and ``-'' indicate the phase of the field. Right two panels: $|E_y|^2$ for the superposition of two modes $|E_{y1} \pm E_{y2}|^2$. (d-f) Calculated dependence $k_z(k_x)$ for the periodic array at three representative wavelengths $\lambda$ marked in panel (b) by the horizontal dashed lines.} 
\label{fig:dispersion}
\end{figure}

To achieve the desired functionality and support our qualitative reasoning, we design subwavelength structured waveguides etched from crystalline silicon on sapphire substrate as shown in Figure~\ref{fig:dispersion}a. Such nanostructuring of the waveguides provides the periodic permittivity modulation essential for dispersion engineering of the two modes~\cite{ChebenNature2018}. In our case, by tuning the width of the gaps cut in an initially homogeneous silicon ridge waveguide, we tailor the dispersion of the two modes of interest~-- a fundamental hybrid HE$_{11}$ mode with dominant $x$ component of the magnetic field and the first quasi-TE$_{01}$ mode with dominant $z$ component of the magnetic field along the waveguide axis. Substantially different strengths of the $z$-component of the electric field for these modes result in different dispersion shifts as we vary the gap width~\cite{JohnsonPRE2002}. By optimizing the geometric parameters, we achieve an intersection of the dispersion branches of these modes in the telecom range $\lambda \approx 1.55\;\mu m$, as shown by the green curves in Figure~\ref{fig:dispersion}b (see Supplementary Materials for details). The respective distributions of electric field for those modes are shown in Figure~\ref{fig:dispersion}c.


Performing full-wave numerical simulation of the finite array, we recover the bands of waveguide modes shown by shaded areas in Figure~\ref{fig:dispersion}b. The blue areas correspond to the bands formed by the two modes of interest with dominant polarization of the magnetic field in the plane of the substrate, while the grey area corresponds to an additional mode with dominant out-of-plane polarization of the magnetic field. Due to the presence of the substrate, the modes with different polarizations mix with each other, becoming hardly distinguishable at long wavelengths $\lambda>1600$~nm.

To access the topological properties of the designed structure, we examine the dependence of the propagation constant $k_z$ on Bloch wave number $k_x$. If the wavelength of operation $\lambda$ is close to the target value 1.55~$\mu m$, the spectrum of the propagation constants features a gap and the associated topological invariant -- Zak phase -- acquires a nonzero value (see Supplementary Materials for further details and analytical tight-binding model). Moreover, even in the absence of perfect mode degeneracy, the system retains its topological nature, up to a critical detuning between the propagation constants $k_z$ of the two modes [see green lines in Figure~\ref{fig:dispersion}b], while the magnitude of the critical detuning depends on the coupling strength between the neighboring  waveguides. 


Figures~\ref{fig:dispersion}d-f show that the gap in the spectrum of propagation constants is close to maximal at wavelength $\lambda=1555$~nm when the dispersion curves of the two modes intersect. At the same time, for wavelengths $\lambda=1528$~nm or $\lambda=1610$~nm, at which the detuning in $k_z$ between the two modes becomes large enough, the topological gap closes and the system transitions to a trivial phase; hence the edge states disappear. Simulating a finite array of 7 waveguides numerically, we indeed reveal the edge states with the dispersion shown in Figure~\ref{fig:dispersion}b by the red curves. The edge states arise within approximately $50$~nm wavelength interval. Hence, by sweeping the wavelength of operation, one can readily adjust the width of the gap in the spectrum of propagation constants $k_z$ and control the localization  of the topological mode, switching it from tight localization at the edge to complete delocalization over the entire array.





An interesting feature of the predicted topological modes evident already from our qualitative analysis is their hybrid nature: they are the superposition of symmetric ${\bf E}_1$ and antisymmetric ${\bf E}_2$ modes of the individual waveguides, $\mathbf{E}_{{\rm edge}} \approx \mathbf{E}_{1} + \mathbf{E}_{2}$ (see Supplementary Materials for details and convention on the phase choice). As a result, the near field of the edge modes is strongly asymmetric and it almost vanishes in the direction towards the bulk of the waveguide lattice [Figure~\ref{fig:dispersion}c] which effectively decouples the edge state from the rest of the array. In contrast, the field of the bulk states characterized by a $\pi$ phase difference between the waveguide modes, $\mathbf{E}_{{\rm bulk}} \approx \mathbf{E}_{1} - \mathbf{E}_{2}$, is strongly suppressed at the edge of the array being coupled to the rest of the lattice. 


The hybrid nature of the topological mode provides a useful tool for its coherent control. Since the fields of bulk and edge modes are strongly asymmetric relative to the center of the edge waveguide, they can be selectively excited by launching the excitation through the auxiliary homogeneous single-mode bus waveguide (SMWG) placed either from the left or from the right edge from the edge waveguide (Fig.~\ref{fig:scheme}a). Here, the width of the SMWG is adjusted in such a way that the dispersion of its fundamental mode crosses the intersection point of the two modes of the multimode waveguide.

\begin{figure}[t!]
\centering
\includegraphics[width=0.6\textwidth]{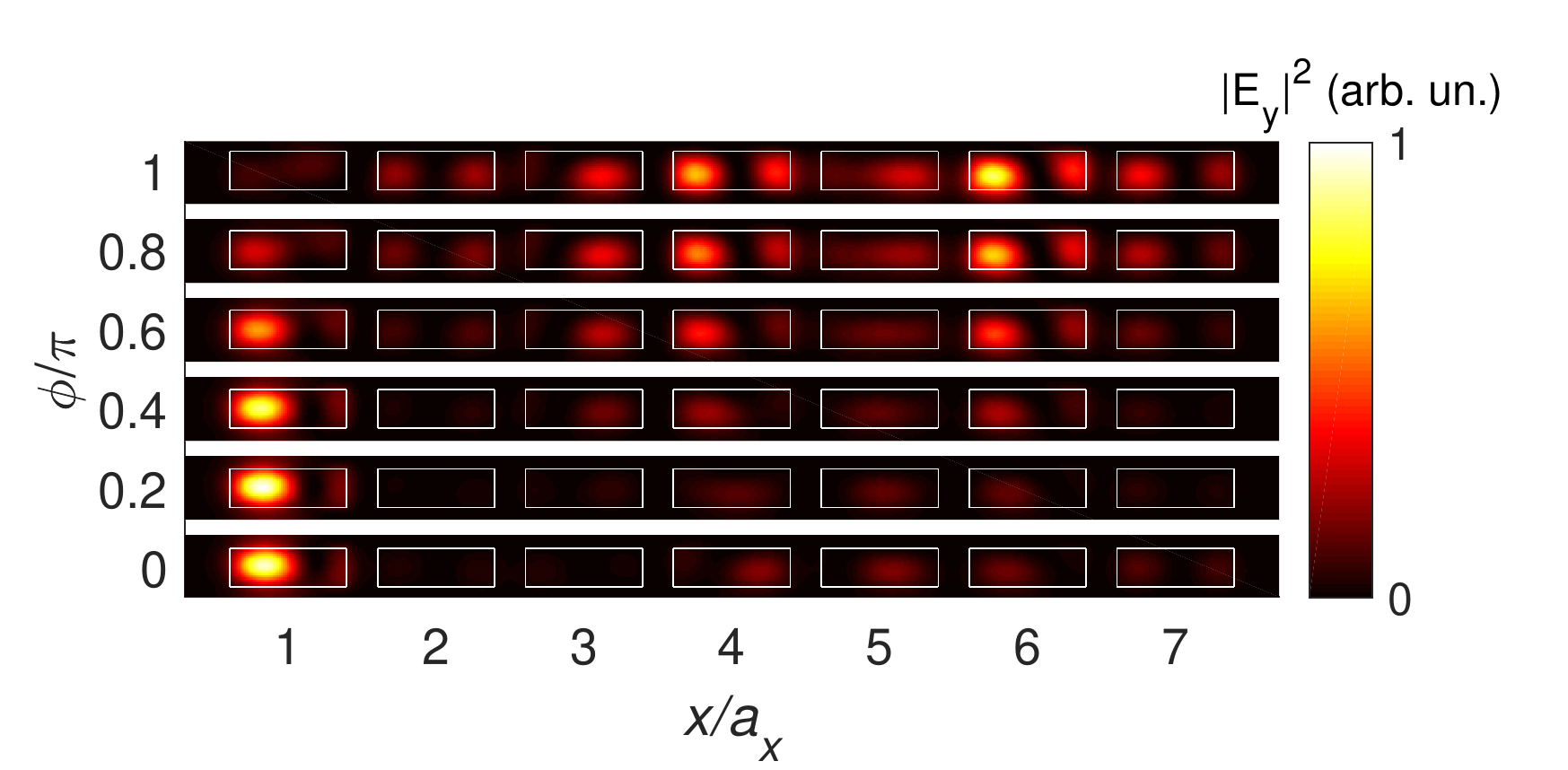}
\caption{Distribution of the field $|E_y|^2 = \left|E_{r}\,\cos\frac{\phi}{2}  + i\,E_{l}\,\sin \frac{\phi}{2}\right|^2$ in the array of 7 waveguides at the output unit cell for the different values of $\phi$. $E_{r,l}$ denote the electric fields excited in the waveguide array via right or left directional coupler, respectively.}
\label{fig:phase}
\end{figure}


To quantify the degree of coherent control of our system, we simulate the field distribution at the output facet of the array for the various relative phases $\phi$ between the two waveguide modes launched into the edge waveguide. To achieve that, we assume that the left and right single-mode bus waveguides are excited simultaneously with $\pi/2$ phase shift between the excitations: $\VB{E}_{tot} = \VB{E}_r\,\cos \frac{\phi}{2}  + i\,\VB{E}_l\,\sin \frac{\phi}{2}$. Since excitations of right and left couplers translate into ${\bf E}_1+{\bf E}_2$ and ${\bf E}_1-{\bf E}_2$ superpositions of the waveguide modes, the output field reads $\VB{E}_{tot}\propto \VB{E}_1\,e^{i\phi/2}+\VB{E}_2\,e^{-i\phi/2}$ which is indeed a superposition of the two waveguide modes with $\phi$ phase difference between them. Numerical results presented in Figure~\ref{fig:phase} confirm our expectation showing efficient localization of light for in-phase excitation of $\VB{E}_1$ and $\VB{E}_2$ and almost complete delocalization for the out-of-phase excitation.


\section{Results}

\begin{figure}[t!]
\centering
\includegraphics[width=1\textwidth]{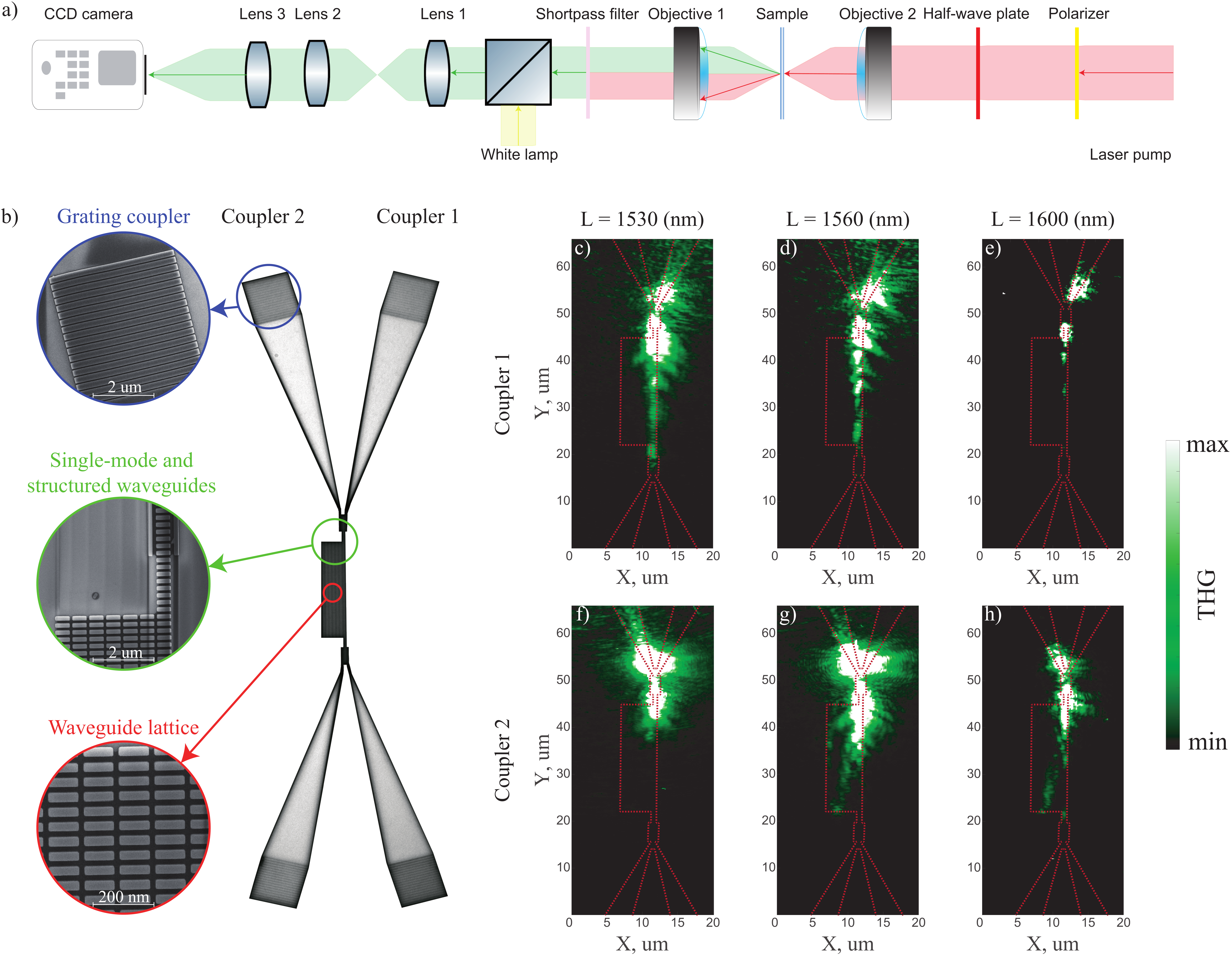}
\caption{(a) A scheme of the experimental setup for third harmonic measurements in the transmission geometry. Red arrow indicates the excitation laser path, green arrow - third harmonic signal, yellow arrow - white lamp for the visualization of the sample. Linearly polarized light is focused on the grating coupler through the objective 2, the third harmonic signal generated in the sample is collected through the objective 1. After filtering with a shortpass filter to remove the remnants of the fundamental frequency, the sample is imaged with a CCD camera. (b) Optical image of the photonic structure and close-up SEM images of its main components: the grating coupler (circled with blue), single-mode and structured waveguides (circled with green), coupled waveguide lattice (circled with red). (c-h) Propagation of topological edge modes and bulk modes for different excitation wavelengths visualized via third harmonic signal. (c-e) Excitation through coupler 1 (edge mode regime). (f-h) Excitation through coupler 2 (bulk mode regime). Red dashed lines in (c-h) show the contours of the structure.} 
\label{fig:exp_res}
\end{figure}

The designed array of nano-waveguides was fabricated on a silicon on sapphire substrate with a $600$~nm device layer using standard CMOS-compatible fabrication techniques (see  Methods section). The optical image of one of the fabricated devices, as well as zoom-in SEM images of its elements, are provided in Figure~\ref{fig:exp_res}b. To excite the structure from the far field, we have added linear tapers and fully etched grating couplers to the designed SMWGs.



As a convenient tool to visualize the field of bulk and edge states, we exploit the third-harmonic generation (THG) process. Since the topological states are highly confined and the third harmonic signal is proportional to the cube of the field at the fundamental frequency, this technique provides excellent spatial resolution to capture the finest details of the field distribution~\cite{Kruk-Poddubny,Smirnova-PRL}  allowing to outcouple light from the waveguide structure for far-field detection.

The scheme of the experimental setup is shown in Figure~\ref{fig:exp_res}a. Linearly polarized laser pulses with magnetic field oriented along the grating bars (TM-polarization) are focused on one of the grating couplers marked by blue in Figure~\ref{fig:exp_res}b from the substrate side. The pulses are generated by an optical parametric amplifier (see Methods for details), which allows us to tune the excitation wavelength within a broad spectral range. Light coupled to the planar  waveguide via the grating passes through the taper region and excites one of the SMWGs. The energy from the SMWG is further transferred to the extended structured waveguide (marked by green in Figure~\ref{fig:exp_res}b) in the form of in-phase or out-of-phase superposition of the two waveguide modes depending on the chosen excitation port: right (1) or left (2) coupler. The area of the sample containing the directional coupler and the waveguide array is then imaged on a CCD camera. The detected third harmonic signal allows us to visualize the propagation of the modes in the waveguide array.

In agreement with our simulations, the coupling of light through coupler 1 leads to the excitation of a topological edge mode, which manifests itself as a strong THG signal localized at the edge of the waveguide array (Figure~\ref{fig:exp_res}c-e). In contrast, excitation through the coupler 2 reverses the phase difference between the modes of the extended waveguide, which causes the diffraction of waves into the bulk of the array (Figure~\ref{fig:exp_res}f-h). The topological edge mode emerges in the range of wavelengths from 1530 nm to 1600 nm, consistent with our numerical results; the localization of the edge mode at a wavelength of 1560 nm appears to be especially pronounced (Figure~\ref{fig:exp_res}d), which indicates the optimal combination of the directional coupler efficiency and the proximity to the degeneracy point of the two waveguide modes. Notably, the third harmonic signal from the waveguide array decreases with the increase of the detuning between the propagation constants of the waveguide, which is mostly due to the limited bandwidth of the directional coupler. Even larger detunings of the excitation from the degeneracy point lead to the closure of the topological gap. In this regime, bulk modes are excited both when the structure is fed through coupler 1 or coupler 2. This trivial case is illustrated in Supplementary Materials.

These experimental results confirm the emergence of  topological edge state mediated by the accidental degeneracy of the modes of the individual waveguide. Hybrid nature of the edge state enables its coherent control and the possibility to continuously tune the operation of the structure from bulk to edge modes. At the same time, the degree of the edge mode localization is controlled by the excitation wavelength. Thus, the designed setup uncovers novel possibilities of integrated photonic topological structures extending the range of their possible functionalities.


\section{Conclusions}\label{sec:outro}

In conclusion, our experiments demonstrate a novel avenue to design highly controllable topological insulator structures and meta-devices at the nanoscale. Owing to the multimode nature of the lattice, the observed topological modes have hybrid nature exhibiting different excitation efficiencies for the different input ports of the optical signal. This feature gives an immediate access to their coherent control and switching in real-time. 

At the same time, the topological properties of our structure can be manipulated by changing the wavelength of excitation which determines not only the localization length, but also the very existence of the topological edge mode. We believe that the proposed strategy could be also beneficial for other classes of topological meta-devices such as topological waveguides, cavities and lasers.


\section{METHODS} \label{sec:Methods} %
\textbf{Sample Preparation.}
Designed nanostructures were fabricated using standard CMOS-compatible techniques. A silicon on sapphire substrate with a 600 nm device layer was purchased from \textit{Universitywafers}. The substrate was cleaned in an acetone ultrasonic bath followed by isopropanol rinse, 5 min O2 plasma descum, and 10 min hot plate degassing at  $180^\circ C $. 230 nm HSQ mask was spin-coated on the surface (60 s at 5000 rpm) and baked on a hot plate for 2 minutes at $85^\circ C $. Resist was patterned using \textit{JEOL} 100 KeV electron-beam lithography system with 1 nA current. To minimize the charging effect of the insulating substrate, a 10 nm conductive polymer layer (\textit{Electra 92} from \textit{Allresist}) was deposited on the surface before the lithography. After the exposure, the conductive polymer was removed using DI water, followed by a 2 min post-exposure bake at $195^\circ C $. Resist was developed in TMAH (25\%) for 70 s followed by 60 s DI water- and 20 s isopropanol-rinsing. The pattern was transferred to the Si slab using the ICP RIE (\textit{Oxford PlasmaPro System 100 Cobra}) in HBr(20 sccm)/Ar(10 sccm) gas mixture for 2 min 30 s (ICP 1500 W, RF 40 W, pressure 8 mTorr). A subsequent buffered HF dip removed the residuals of the mask. 

\textbf{Third-harmonic spectroscopy.} To image the propagation of the modes of the structured waveguide array, we excite the sample from the substrate side with a 10x Mitutoyo objective focussing the laser radiation on one of the grating couplers at a small incidence angle. Femtosecond laser pulses are provided by the optical parametric amplifier (Light Conversion Orpheus) pumped by 250 fs Yb laser (LightConversion Pharos) at a repetition rate of 1 MHz. The automated parametric ampification system allows us to tune the excitation wavelength within a broad spectral range.  
The area of the sample containing the directional coupler and the waveguide array was imaged on a CCD camera (PyLoN 400BR\underline{ }eXcelon) with a 50x Mitutoyo objective. The residual scattered light at the fundamental laser frequency is blocked with a shortpass filter.




\subsection{Acknowledgments \& Funding Sources}

We acknowledge Maxim Mazanov and Prof. Rodrigo Vicencio for valuable discussions. Theoretical and numerical studies were supported by Priority 2030 Federal Academic Leadership Program. Optical characterization of the structure was supported by the Ministry of Science and Higher Education of the Russian Federation (project No.~075-15-2021-589). Sample fabrication and initial characterization were supported by the Air Force Office of Scientific Research, the Simons Foundation and the European Union’s Horizon 2020 research and innovation
programme under the Marie Sklodowska-Curie grant agreement N 845287. M.A.G. acknowledges partial support by the Foundation for the Advancement of Theoretical Physics and Mathematics ``Basis".

Supporting Information Available. This material is available free of charge via the Internet at http://pubs.acs.org

\bibliography{TopologicalLib}

\end{document}